\documentclass[10pt,reqno,a4paper]{article}
\usepackage{amsmath,amssymb,dsfont,graphicx,cite,rotating,setspace}
\def\beq{\begin{equation}}
\def\eeq{\end{equation}}
\def\bea{\begin{eqnarray}}
\def\eea{\end{eqnarray}}
\newtheorem{theorem}{Theorem}

\makeatletter
\expandafter\let\expandafter
\reset@font\csname reset@font\endcsname
\def\subeqnarray{\arraycolsep1pt
    \def\@eqnnum\stepcounter##1{\stepcounter{subequation}
        {\reset@font\rm(\theequation\alph{subequation})}}
\jot5mm     \eqnarray}

\makeatother

\def\ri{{\rm{i}}}

\def\tilde{\widetilde}

\def\su2{{\mathfrak {su}}(2)}
\def\e3{{\mathfrak {e}}(3)}

\begin{document}
\medskip

\begin{center}
{\Large \bf {Classification of discrete equations linearizable  by point transformation  on a square lattice}\\[0.5cm]
{ }}

\bigskip

\noindent{\bf  Christian Scimiterna and Decio Levi}

\medskip

\noindent {Dipartimento di Ingegneria Elettronica \\
Universit\`a degli Studi Roma Tre and  INFN Sezione di Roma Tre \\
Via della Vasca Navale 84, 00146 Roma, Italy}

\medskip
\noindent{{\bf E-mail:} scimiterna@fis.uniroma3.it;  levi@roma3.infn.it }
\bigskip

\end{center}

\begin{abstract}

We provide a complete set of linearizability conditions for  nonlinear partial difference equations defined on four points and, using them, we classify all linearizable multilinear partial difference equations defined on four points up to a M\"obious transformation.
\end{abstract}

\medskip

\noindent PACS numbers: 

\noindent Mathematics Subject Classification: 

\medskip
\section{Introduction}
\begin{figure}[htbp] \label{fig1}
\begin{center}
\setlength{\unitlength}{0.1em}
\begin{picture}(200,140)(-50,-20)
 \put( 0,  0){\line(1,0){100}}
  \put( 100,  0){\line(-1,0){100}}
\put( 0,100){\line(1,0){100}}
  \put( 100,100){\line(-1,0){100}}
\put(  0, 0){\line(0,1){100}}
  \put(  0, 100){\line(0,-1){100}}
\put(100, 0){\line(0,1){100}}
  \put(100, 100){\line(0,-1){100}}
   \put(97, -3){$\bullet$}
   \put(-3, -3){$\bullet$}
   \put(-3, 97){$\bullet$}
   \put(97, 97){$\bullet$}
  \put(-32,-13){$u_{n,m}$}
  \put(103,-13){$u_{n+1,m}$}
  \put(103,110){$u_{n+1,m+1}$}
  \put(-32,110){$u_{n,m+1}$}
\end{picture}
\caption{The  quad--graph where a partial difference equation is defined}
\end{center}
\end{figure}
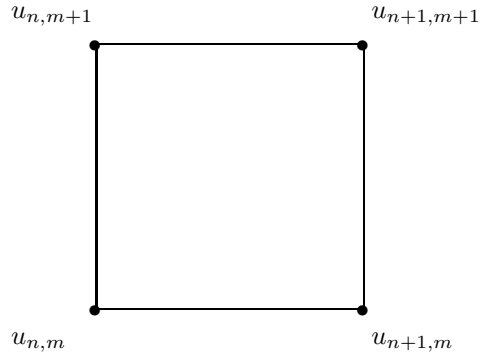

In a series of papers \cite{ls0,ls,ls1,ls2} one has provided necessary conditions for the linearizability of  real dispersive multilinear difference equations on a quad--graph (see Fig. \ref{fig1}).
and on three points (Fig. 2). 
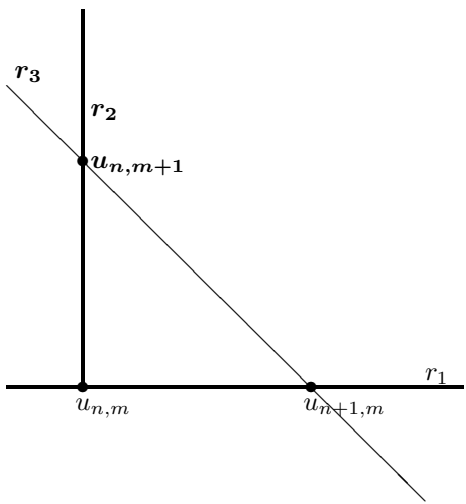
\begin{figure} \label{fig2}
\setlength{\unitlength}{1mm}
\begin{center}
\begin{picture}(100,51)
 \linethickness{1.pt}
     \put(80,10) {\circle*{1,5}}
  \put(50,10) {\circle*{1,5}}
   \put(50,40) {\circle*{1,5}}
\put(95,11){$r_1$}
\put(51,46){\boldmath $r_2$}
\put(41,51){\boldmath $r_3$}
\put(49,07){$u_{n,m}$}
\put(79,07){$u_{n+1,m}$}
\put(51,39){\boldmath $u_{n,m+1}$}
\put(40,50){\line(1,-1){55}}
\put(40,10) {\line(1,0){60}}
\put(50,10) {\line(0,1){50}}

\end{picture}
\end{center}
\caption[]{Points related by an equation defined on three points. }
\end{figure}

These conditions, obtained by considering the existence of point transformations and symmetries have been sufficient to classify the multilinear equations defined on three points \cite{ls1} but not those defined on four points. In \cite{ls2} we considered the problem from the point of view of the symmetries, both point and nonlocal. In this way we get a different set of conditions with respect to those obtained before which, however, are not yet sufficient to classify the multilinear equations defined on four points. 
So here, using the experience of \cite{ls1} and \cite{ls2} we  construct the largest possible set of linearizability conditions and, through them, we classify the multilinear equations on a square lattice. We assume a   partial difference equation on a quad--graph to be given by
\bea \label{1.1a}
\mathcal E =\mathcal E\left(u_{n,m}, u_{n,m+1}, u_{n+1,m},u_{n+1,m+1}\right)=0, \quad \frac{\partial \mathcal E}{\partial u_{n+i,m+j}}\not=0, \; i,j=0,1,
\eea
for a field $u_{n,m}$ which linearizes into a linear autonomous equation for $\tilde u_{n,m}$
\bea \label{1.3a}
a{\tilde u}_{n,m}+b{\tilde u}_{n+1,m}+c{\tilde u}_{n,m+1}+d{\tilde u}_{n+1,m+1}+e = 0
\eea
with $a$, $b$, $c$, $d$ and $e$ being ($n,m$)--independent arbitrary non zero complex coefficients. The choice that (\ref{1.3a}) be autonomous  is a  restriction but it is also a natural simplifying ansatz when one is dealing with autonomous equations. Moreover, as (\ref{1.1a}, \ref{1.3a}) are taken to be autonomous  equations, i.e. they have no $n,m$ dependent coefficients, they are translationally invariant under shifts in $n$ and $m$. So we can with no loss of generality choose as reference point $n=0$ and $m=0$. This will also be assumed to be  true for  the linearizing point tranformation. 

By a {\it linearizing  point transformation} we mean a transformation
\bea \label{1.3b}
\tilde u_{0,0} = f(u_{0,0})
\eea
between (\ref{1.3a}) and (\ref{1.1a})  characterized by a function depending just from the function $u_{0,0}$ and on some constant parameters. It will be a {\it Lie point transformation} if   $f=f_{0,0}$ satisfies all Lie group axioms and, in particular, the composition law.   In the following we will require much less, i.e. we will only assume the differentiability of the function $f$ up to at least second order.

In Section 2 we  discuss  point transformations, present the linearizability conditions which ensure  that the given equation is  linearizable  and the differential equations which define the transformation $f$. 
In Section 3 we   classify all multilinear equations which belong to the class   \eqref{1.1a} up to a M\"obious transformation while in the final Section we  present  some conclusive remarks and open problems.

\section{Discrete equations defined on a square linearizable  by a point transformation.}
In the autonomous case a generic partial difference equation  \eqref{1.1a}  for the complex function $u_{n,m}\doteq u_{0,0}$ can be rewritten as
\bea
\mathcal E\left(u_{0,0},u_{1,0},u_{0,1},u_{1,1}\right)=0, \quad \frac{\partial \mathcal E}{\partial u_{i,j}}\not=0, \; i,j=0,1,
\label{Roma}
\eea
 We will assume that we can solve (\ref{Roma}) with respect to each one of the four variables in its argument
\begin{subequations}\label{Zero}
\bea \label{Zeroa}
u_{1,1}&=&F\left(u_{0,0},u_{1,0},u_{0,1}\right),\ \ \ F_{,u_{0,0}}\not=0\ \ \ F_{,u_{1,0}}\not=0,\ \ \ F_{,u_{0,1}}\not=0,\,\\ \label{Zerob}
u_{1,0}&=&G\left(u_{0,0},u_{0,1},u_{1,1}\right),\ \ \ G_{,u_{0,0}}\not=0,\ \ \ G_{,u_{0,1}}\not=0,\ \ \ G_{,u_{1,1}}\not=0,\\ \label{Zeroc}
u_{0,1}&=&S\left(u_{0,0},u_{1,0},u_{1,1}\right),\ \ \ S_{,u_{0,0}}\not=0,\ \ \ S_{,u_{1,0}}\not=0\ \ \ S_{,u_{1,1}}\not=0,\\ \label{Zerod}
u_{0,0}&=&T\left(u_{1,0},u_{0,1},u_{1,1}\right),\ \ \ T_{,u_{1,0}}\not=0,\ \ \ T_{,u_{0,1}}\not=0\ \ \ T_{,u_{1,1}}\not=0, 
\eea
\end{subequations}
 and that \eqref{Roma} can be linearized by the linearizing autonomous point transformation \eqref{1.3b} into the linear equation \eqref{1.3a} for the complex function $u_{n,m}\doteq u_{0,0}$, i.e
\bea
a\tilde u_{0,0}+b\tilde u_{1,0}+c\tilde u_{0,1}+d\tilde u_{1,1}+e=0.\label{Janus}
\eea
 Hence, assuming we can solve (\ref{Roma}) with respect to $u_{1,1}$, we can choose as independent variables $u_{0,0}$, $u_{1,0}$ and $u_{0,1}$ and we will have that
\bea
af_{0,0}+bf_{1,0}+cf_{0,1}+df_{1,1}\vert_{u_{1,1}=F}+e=0,\label{Amor}
\eea
\noindent must be identically satisfied  for any $ u_{0,0}$, $u_{1,0}$ and $u_{0,1}$. Differentiating (\ref{Amor}) with respect to $u_{0,0}$, $u_{1,0}$ or $u_{0,1}$, we  obtain
\begin{subequations}\label{Saturno}
\bea
a\frac{df_{0,0}}{du_{0,0}}+d\frac{df_{1,1}}{du_{1,1}}\vert_{u_{1,1}=F}F_{,u_{0,0}}=0,\label{Saturno1}\\ \label{Saturno2}
b\frac{df_{1,0}}{du_{1,0}}+d\frac{df_{1,1}}{du_{1,1}}\vert_{u_{1,1}=F}F_{,u_{1,0}}=0,\\ \label{Saturno3}
c\frac{df_{0,1}}{du_{0,1}}+d\frac{df_{1,1}}{du_{1,1}}\vert_{u_{1,1}=F}F_{,u_{0,1}}=0,
\eea
\end{subequations}
\noindent which have to be identically satisfied for any $ u_{0,0}$, $u_{1,0}$ and $u_{0,1}$. From them, considering that $\frac{df\left(x\right)}{dx}\not=0$, we derive that $d\not=0$, otherwise $a=b=c=e=0$. As a consequence, considering that also  $F_{,u_{0,0}}\not=0$, $F_{,u_{1,0}}\not=0$ and $F_{,u_{0,1}}\not=0$, we have  $a\not=0$, $b\not=0$ and $c\not=0$. Then in all generality we can divide (\ref{Janus}) by $d$ and, introducing the new parameters $\alpha\doteq a/d\not=0$, $\beta\doteq b/d\not=0$, $\gamma\doteq c/d\not=0$ and $\epsilon\doteq e/a$,  (\ref{Janus}) can be rewritten as
\bea
\alpha\tilde u_{0,0}+\beta\tilde u_{1,0}+\gamma\tilde u_{0,1}+\tilde u_{1,1}+\epsilon=0.
\eea
\noindent Defining $\frac{df\left(x\right)}{dx}\doteq H\left(x\right)$, from (\ref{Saturno}) we obtain
\begin{subequations}\label{Isis}
\bea
\frac{F_{,u_{0,0}}}{F_{,u_{1,0}}}=\frac{\alpha H\left(u_{0,0}\right)}{\beta H\left(u_{1,0}\right)},\\
\frac{F_{,u_{0,0}}}{F_{,u_{0,1}}}=\frac{\alpha H\left(u_{0,0}\right)}{\gamma H\left(u_{0,1}\right)}.
\eea
\end{subequations}
From (\ref{Isis}) we get the following linearizability conditions
\begin{subequations}\label{Osiris}
\bea
A\left(x,u_{0,1}\right)\doteq\frac{F_{,u_{0,0}}}{F_{,u_{1,0}}}\vert_{u_{0,0}=u_{1,0}=x}=\frac{\alpha}{\beta},\ \ \ \forall x,\ u_{0,1},\label{Osiris1}\\
B\left(x,u_{1,0}\right)\doteq\frac{F_{,u_{0,0}}}{F_{,u_{0,1}}}\vert_{u_{0,0}=u_{0,1}=x}=\frac{\alpha}{\gamma},\ \ \ \forall x,\ u_{1,0},\label{Osiris2}\\
C\left(x,u_{0,0}\right)\doteq\frac{F_{,u_{0,1}}}{F_{,u_{1,0}}}\vert_{u_{1,0}=u_{0,1}=x}=\frac{\gamma}{\beta},\ \ \ \forall x,\ u_{0,0},\label{Osiris3}\\
\frac{\partial}{\partial u_{0,1}}\frac{F_{,u_{0,0}}}{F_{,u_{1,0}}}=0,\ \ \ \forall u_{0,0},\ u_{1,0},\ u_{0,1},\label{Osiris4}\ \ \ \ \ \ \ \ \\
\frac{\partial}{\partial u_{1,0}}\frac{F_{,u_{0,0}}}{F_{,u_{0,1}}}=0,\ \ \ \forall u_{0,0},\ u_{1,0},\ u_{0,1},\label{Osiris5}\ \ \ \ \ \ \ \ \\
\frac{\partial}{\partial u_{0,0}}\frac{F_{,u_{0,1}}}{F_{,u_{1,0}}}=0,\ \ \ \forall u_{0,0},\ u_{1,0},\ u_{0,1}.\label{Osiris6}\ \ \ \ \ \ \ \ 
\eea
\end{subequations}
Alternatively the conditions (\ref{Osiris1}-\ref{Osiris3}) can be substituted  by the following ones
\begin{subequations}
\bea
\frac{d}{dx}A\left(x,u_{0,1}\right)=0,\ \ \ \forall x,\ u_{0,1},\\
\frac{d}{dx}B\left(x,u_{1,0}\right)=0,\ \ \ \forall x,\ u_{1,0},\\
\frac{d}{dx}C\left(x,u_{0,0}\right)=0,\ \ \ \forall x,\ u_{0,0}.
\eea
\end{subequations}
Taking the (principal value of the) logarithm of (\ref{Saturno1}), we have
\bea
\log\frac{df_{0,0}}{du_{0,0}}-\log\frac{df_{1,1}}{du_{1,1}}\vert_{u_{1,1}=F}=\log\left(-\frac{F_{,u_{0,0}}}{\alpha}\right)\ \left(\mbox{mod}\,2\pi\ri\right).\label{Sat}
\eea
\noindent Then, let us  introduce the linear operator $\mathcal B$ 
\bea \label{lob}
\mathcal B\doteq\frac{\partial}{\partial u_{0,0}}-\frac{F_{,u_{0,0}}}{F_{,u_{1,0}}}\frac{\partial}{\partial u_{1,0}},
\eea
 such that and $B\phi\left(F\left(u_{0,0},u_{1,0},u_{0,1}\right)\right)=0$, where $\phi$ is an arbitrary functions of its argument. When we  apply \eqref{lob}  to (\ref{Sat}), we  obtain an ordinary differential equation describing the evolution of the linearizing transformation
\bea
\frac{d}{d u_{0,0}}\log\frac{d f_{0,0}}{d u_{0,0}}=\frac{1}{F_{,u_{0,0}}F_{,u_{1,0}}}W_{(u_{0,0})}\left[F_{,u_{1,0}};F_{,u_{0,0}}\right],\label{Romulius}
\eea
where $W_{(x)}\left[f;g\right]\doteq fg_{,x}-gf_{,x}$ stands for the Wronskian of the functions $f$ and $g$. Let's remark that the linearizability conditions (\ref{Osiris4}-\ref{Osiris6}) imply that the right hand member of the equation (\ref{Romulius}) does not depend on $u_{1,0}$ and $u_{0,1}$. These conditions were considered in \cite{ls} and had not been sufficient to classify \eqref{1.1a}. Other similar conditions can be obtained starting from \eqref{Saturno2} or \eqref{Saturno3}. 
The linearizability conditions presented here  have  been obtained starting from \eqref{Zeroa}. Similar results could be obtained starting from \eqref{Zerob}, \eqref{Zeroc} or \eqref{Zerod}. However these results would not have provided any really new linearizability condition.
So, here, in the next Section  we start the classifying process from the more basic equations \eqref{Osiris} as we did in the case of equations depending on just three points \cite{ls2}.

\section{Classification of complex autonomous multilinear partial difference equations defined on four points linearizable by a point transformation.}

\noindent Let $\mathcal E\left(u_{0,0},u_{1,0},u_{0,1},u_{1,1}\right)=0$ be the  complex multilinear equation
\bea\label{Mars}
a_{1}u_{0,0}&+&a_{2}u_{1,0}+a_{3}u_{0,1}+a_{4}u_{1,1}+a_{5}u_{0,0}u_{1,0}+a_{6}u_{0,0}u_{0,1}+\\
\nonumber &+&a_{7}u_{0,0}u_{1,1}+a_{8}u_{1,0}u_{0,1}+a_{9}u_{1,0}u_{1,1}+a_{10}u_{0,1}u_{1,1}+\\
\nonumber &+&a_{11}u_{0,0}u_{1,0}u_{0,1}+a_{12}u_{0,0}u_{1,0}u_{1,1}+a_{13}u_{0,0}u_{0,1}u_{1,1}+\\
\nonumber &+&a_{14}u_{1,0}u_{0,1}u_{1,1}+a_{15}u_{0,0}u_{1,0}u_{0,1}u_{1,1}+a_{16}=0,
\eea
\noindent where $a_{j}$, $j=1,\ldots$, $16$, are arbitrary complex free parameters. This equation is invariant under a M\"obious transformation of the dependent variable
\bea
u_{0,0}\doteq\frac{b_{1}v_{0,0}+b_{2}}{b_{3}v_{0,0}+b_{4}},
\eea
where $b_{k}$, $k=1,\ldots$, $4$ are four arbitrary complex parameters such that $b_{1}b_{4}-b_{2}b_{3}\not=0$. As we operate in the field of complex numbers and we classify up to M\"obious transformations, using inversions, dilations and translations we can always simplify \eqref{Mars} by setting either  {\bf 1)} $a_{15}=a_{16}=0$ or {\bf 2)} $\sum_{k=1}^{4}a_{k}=\sum_{k=5}^{10}a_{k}=\sum_{k=11}^{14}a_{k}=a_{15}=0$, $a_{16}=1$. Let's now apply to these two cases the six necessary linearizability conditions (\ref{Osiris}). This amounts to solving a system of 96 algebraic in general nonlinear equations involving the coefficients $a_{j}$, $j=1,\ldots$, $14$. Their solution implies, through the integration of the differential equation (\ref{Romulius}), that the function $f\left(x\right)$ appearing in the linearizing transformation (\ref{1.3a}) can only be of the following two types:
\begin{enumerate}
\item The fractional linear function
\bea
f\left(x\right)=\frac{c_{1}x+c_{2}}{c_{3}x+c_{4}},\ \ \ c_{1}c_{4}-c_{2}c_{3}\not=0, 
\eea
which, as the classification is up to M\"obious transformations of $u_{0,0}$, can always be reduced to be $f\left(x\right)=1/x$;
\item The (principal branch) of the logarithmic function
\bea
f\left(x\right)=d_{1}\log\left(\frac{d_{2}x+d_{3}}{d_{4}x+d_{5}}\right)+d_{6}\ \ \ d_{1}\not=0,\ \ \ d_{2}d_{5}-d_{3}d_{4}\not=0,
\eea
which, as the classification is up to M\"obious transformations, can always be reduced 
 to  $f\left(x\right)=\log\left(x\right)$. Moreover in this case  the ratios $\alpha/\beta$ and $\alpha/\gamma$ are always real and of modulus $1$, so that the possible linear equations can be only of the following four types:
\begin{subequations}
\bea
\alpha\left(\tilde u_{0,0}+\tilde u_{1,0}+\tilde u_{0,1}\right)+\tilde u_{1,1}+\epsilon=0;\\
\alpha\left(\tilde u_{0,0}+\tilde u_{1,0}-\tilde u_{0,1}\right)+\tilde u_{1,1}+\epsilon=0;\\
\alpha\left(\tilde u_{0,0}-\tilde u_{1,0}+\tilde u_{0,1}\right)+\tilde u_{1,1}+\epsilon=0;\\
\alpha\left(\tilde u_{0,0}-\tilde u_{1,0}-\tilde u_{0,1}\right)+\tilde u_{1,1}+\epsilon=0.
\eea
\end{subequations}
It is easy to prove that, if  the transformation $\tilde u_{0,0}\doteq\log\left(u_{0,0}\right)$ has to produce a multilinear equation for $u_{0,0}$, we must have $\alpha=\pm 1$. In fact, as $F\left(u_{0,0},u_{1,0},u_{0,1}\right)$ should be a fractional linear function of $u_{0,0}$ with coefficients depending on $u_{1,0}$ and $u_{0,1}$, we have that the relations 
\bea \label{new}
u_{0,0}^{\alpha}\left(e_{1}u_{0,0}+e_{2}\right)=e_{3}u_{0,0}+e_{4},\ \ \ e_{j}=e_{j}\left(u_{1,0},u_{0,1}\right),\ \ \ j=1,\ldots,4,\nonumber
\eea
where $e_{1}e_{4}-e_{2}e_{3}$ is not identically zero for all $ u_{1,0}$ and $u_{0,1}$, must be identically satisfied for all $ u_{0,0}$. Differentiating \eqref{new} twice with respect to $u_{0,0}$, we get $e_{1}\left(\alpha+1\right)u_{0,0}+e_{2}\left(\alpha-1\right)=0$ identically for all $ u_{0,0}$, so that
\bea
e_{1}\left(\alpha+1\right)=0,\ \ \ e_{2}\left(\alpha-1\right)=0.\nonumber
\eea
Considering that $e_{1}$ and $e_{2}$ cannot be simultaneously identically zero, it follows that $\alpha=\pm 1$. In this way we  obtain a set of eight linear equations  corresponding to eight linearizable nonlinear equations. 
\end{enumerate}
 Hence we can summarize the results obtained in  the following Theorem:
\begin{theorem} \label{t1} Apart from the class of equations linearizable by a M\"obious transformation, which can be represented up to a M\"obious transformation of the dependent variable (eventually composed with an exchange of the independent variables $n\leftrightarrow m$) by the equation
\bea
w_{0,1}w_{1,1}\left(w_{1,0}+\beta w_{0,0}\right)+w_{0,0}w_{1,0}\left(\gamma w_{1,1}+\delta w_{0,1}\right)+\epsilon w_{0,0}w_{1,0}w_{0,1}w_{1,1}=0,\ \ \ \epsilon=0,1,\ \ \ \beta,\gamma,\delta\not=0,
\eea
linearizable by the inversion $\tilde u_{0,0}=1/w_{0,0}$ to the equation
\bea
\tilde u_{0,0}+\beta\tilde u_{1,0}+\gamma\tilde u_{0,1}+\delta\tilde u_{1,1}+\epsilon=0,
\eea
the only other linearizable equations are up to a M\"obious transformation of the dependent variable (eventually composed with an exchange of the independent variables $n\leftrightarrow m$), represented by the following six nonlinear equations
\begin{subequations}
\bea
&w_{0,0}w_{1,0}w_{0,1}w_{1,1}-1&=0,\label{Maro1}\\
&w_{0,0}-w_{1,0}w_{0,1}w_{1,1}&=0,\label{Maro2}\\
&w_{1,0}-w_{0,0}w_{0,1}w_{1,1}&=0,\label{Maro3}\\
&w_{1,1}-w_{0,0}w_{1,0}w_{0,1}&=0,\label{Maro4}\\
&w_{0,1}w_{1,1}-\theta w_{0,0}w_{1,0}&=0,\label{Maro5}\\
&w_{0,0}w_{1,1}-\theta w_{1,0}w_{0,1}&=0,\label{Maro6}
\eea
\noindent where $\theta\not=0$ is an otherwise arbitrary complex parameter. They are linearizable by the transformation $\tilde u_{0,0}=\log w_{0,0}$ to the equations
\bea
&\tilde u_{0,0}+\tilde u_{1,0}+\tilde u_{0,1}+\tilde u_{1,1}&=2\pi\ri z,\label{Lati1}\\
&-\tilde u_{0,0}+\tilde u_{1,0}+\tilde u_{0,1}+\tilde u_{1,1}&=2\pi\ri z,\label{Lati2}\\
&\tilde u_{0,0}-\tilde u_{1,0}+\tilde u_{0,1}+\tilde u_{1,1}&=2\pi\ri z,\label{Lati3}\\
&-\tilde u_{0,0}-\tilde u_{1,0}-\tilde u_{0,1}+\tilde u_{1,1}&=2\pi\ri z,\label{Lati4}\\
&-\tilde u_{0,0}-\tilde u_{1,0}+\tilde u_{0,1}+\tilde u_{1,1}&=\log\theta+2\pi\ri z,\label{Lati5}\\
&\tilde u_{0,0}-\tilde u_{1,0}-\tilde u_{0,1}+\tilde u_{1,1}&=\log\theta+2\pi\ri z,\label{Lati6}
\eea
\end{subequations}
where $\log$ always stands for the principal branch of the complex logarithmic function and where $\log\theta$ stands for the principal branch of the complex logarithmic function of the parameter $\theta$.
\end{theorem}
Let's note that, given a solution $w_{0,0}$ of eqs. (\ref{Maro1}-\ref{Maro6}), the choice of the principal branch of the complex logarithm function in the transformation $\tilde u_{0,0}=\log w_{0,0}$ reflects in the inhomogeneous terms of the linear eqs. (\ref{Lati1}-\ref{Lati6}) and is always $|z|\leq 2$. This can be easily seen considering that we must have $2\pi|z|\leq |Im\left(u_{0,0}\right)|+|Im\left(u_{1,0}\right)|+|Im\left(u_{0,1}\right)|+|Im\left(u_{1,1}\right)|\leq 4\pi$ (with a slight difference in \eqref{Lati5} and \eqref{Lati6}). Through the translation $\tilde u_{0,0}=v_{0,0}+\pi\ri z/2$ the linear equation (\ref{Lati1}) can be made homogeneous. As a consequence in (\ref{Maro1}) $w_{0,0}=e^{v_{0,0}}e^{\pi\ri z/2}$, so that the integer parameter $z$ can take the values \newline $z=-1$, $0$, $1$, $2$. Hence we  conclude that the term $2\pi\ri z$ in (\ref{Lati1}) takes into account the discrete symmetry of (\ref{Maro1}) given by $w_{0,0}\rightarrow\zeta w_{0,0}$, where $\zeta$ stands for one of the four  roots of unity $\zeta=1$, $-1$, $\ri$ and $-\ri$. The same happens  when applying  the translation $\tilde u_{0,0}=v_{0,0}+\pi\ri z$ to the linear equations (\ref{Lati2}, \ref{Lati3}) and $\tilde u_{0,0}=v_{0,0}-\pi\ri z$ to  \eqref{Lati4}. In (\ref{Maro2}, \ref{Maro3}, \ref{Maro4}) we have $w_{0,0}=e^{v_{0,0}}e^{\pm\pi\ri z}$ so that, without any loss of generality, we can restrict the values of $z$ to  $z=0$, $1$. The term $2\pi\ri z$ in (\ref{Lati2}, \ref{Lati3}, \ref{Lati4}) takes into account the discrete symmetry of (\ref{Maro2}, \ref{Maro3}, \ref{Maro4}) given by $w_{0,0}\rightarrow\zeta w_{0,0}$, where $\zeta$ stands for one of the two  roots of unity $\zeta=1$ and $-1$. The same happens for the linear equations (\ref{Lati5}) when we consider  the non autonomous translation $\tilde u_{0,0}=v_{0,0}+\left(\log\theta +2\pi\ri z\right)m/2$. In (\ref{Maro5}) $w_{0,0}=e^{v_{0,0}}e^{\pi\ri mz}\theta^{m/2}$ 
so that, without any loss of generality, we can also restrict the values $z$ to  $z=0$, $1$. Hence we conclude that the term $2\pi\ri z$ in (\ref{Lati5}) takes into account the discrete symmetry of (\ref{Maro5}) given by $w_{0,0}\rightarrow\left(-1\right)^m w_{0,0}$. The same happens for the linear equation (\ref{Lati6}) with the non autonomous translation $\tilde u_{0,0}=v_{0,0}+\left(\log\theta+2\pi\ri z\right)nm$. Now, as $e^{2\pi\ri nmz}=1$  for any $ z$, in (\ref{Maro6}) $w_{0,0}=e^{v_{0,0}}\theta^{nm}$ so that, without any loss of generality, we can choose $z=0$.  
\section{Concluding remarks and outlook}
In this paper we have classified all multilinear partial difference equations which can be linearized by a point transformation. The resulting linearizable equations are presented in Theorem \ref{t1} together with their linearized counterparts. 

It seems interesting at this point to try to analyze the case of more complicate classes of equations involving more lattice points and which could provide in the continuous case parabolic or elliptic partial differential equations. Work on this is in progress.

\section*{Acknowledgements}

 DL  and CS have been partly supported by the
Italian Ministry of Education and Research, 2010 PRIN ``Continuous and discrete
nonlinear integrable evolutions: from water waves to symplectic maps".


\bigskip

\bigskip



\begin{thebibliography}{99}
\bibitem{ls0}C. Scimiterna and D. Levi,  C-Integrability Test for Discrete Equations via Multiple Scale Expansions,	
{\it SIGMA} {\bf 6} (2010), 070, 17 pages,     arXiv:1005.5288,     http://dx.doi.org/10.3842/SIGMA.2010.070.

\bibitem{ls1} D. Levi a and C. Scimiterna, Linearizability of Nonlinear Equations on a Quad-Graph by a Point, Two Points and Generalized Hopf-Cole Transformations, {\it SIGMA}  {\bf 7}  (2011), 079, 24 pages,      arXiv:1108.3648,      http://dx.doi.org/10.3842/SIGMA.2011.079.

\bibitem{ls} D. Levi and C. Scimiterna, Linearization through symmetries for discrete equations, submitted to SIGMA 2012.

\bibitem {ls2} C. Scimiterna and D. Levi, Three points partial difference equations linearizable by local and nonlocal transformations, submitted to J. Phys. A, 2012.
\end{thebibliography}
\end{document}